\documentclass[twocolumn,showpacs,preprintnumbers,amsmath,amssymb,prl,superscriptaddress]{revtex4}

\usepackage{graphicx}
\usepackage{color}
\usepackage{multirow}

\def\SIO{$\rm Sr_2IrO_4$}
\def\NIO{$\rm Na_2IrO_3$}
\def\LIO{$\rm Li_2IrO_3$}
\def\AIO{$\rm {\textit A}_2IrO_3$}

\def\jeff{$j_{\rm{eff}}$}

\begin{document}

\title{Crystal field splitting and correlation effect on the electronic structure of $A_2{\rm IrO}_3$}

\author{H.~Gretarsson}
\author{J. P. Clancy}
\affiliation{Department of Physics, University of Toronto, 60
St.~George St., Toronto, Ontario, M5S 1A7, Canada}
\author{X. Liu}
\author{J.~P. Hill}
\author{Emil Bozin}
\affiliation{CMP\&MS Department, Brookhaven National Laboratory,
Upton, New York 11973, USA}
\author{Yogesh Singh}
\affiliation{Indian Institute of Science Education and Research Mohali, Sector 81, SAS
Nagar, Manauli PO 140306, India }
\author{S. Manni}
\author{P. Gegenwart}
\affiliation{I. Physikalisches Institut, Georg-August-Universit\"at
G\"ottingen, D-37077, G\"ottingen, Germany}
\author{Jungho~Kim}
\author{A. H. Said}
\author{D.~Casa}
\author{T.~Gog}
\author{M.~H. Upton}
\affiliation{Advanced Photon Source, Argonne National Laboratory,
Argonne, Illinois 60439, USA}
\author{Heung-Sik Kim}
\author{J. Yu}
\affiliation{Department of Physics and Astronomy, Seoul National
University, Seoul 151-747, Korea}
\author{Vamshi M. Katukuri}
\author{L. Hozoi}
\author{Jeroen van den Brink}
\affiliation{Institute for Theoretical Solid State Physics, IFW Dresden, Helmholtzstr.~20, 01069 Dresden, Germany}
\author{Young-June Kim}
\email{yjkim@physics.utoronto.ca} \affiliation{Department of
Physics, University of Toronto, 60 St.~George St., Toronto, Ontario,
M5S 1A7, Canada}

\date{\today}

\begin{abstract}
The electronic structure of the honeycomb lattice iridates \NIO\ and \LIO\ has been investigated using resonant inelastic x-ray scattering (RIXS). Crystal-field split $d$--$d$ excitations are resolved in the high-resolution RIXS spectra. In particular, the splitting due to non-cubic crystal fields, derived from the splitting of \jeff=3/2 states, is much smaller than the typical spin-orbit energy scale in iridates, validating the applicability of \jeff\ physics in \AIO. We also find excitonic enhancement of the particle-hole excitation gap around 0.4 eV, indicating that the nearest-neighbor Coulomb interaction could be large. These findings suggest that both \NIO\ and \LIO\ can be described as spin-orbit Mott insulators, similar to the square lattice iridate \SIO.

\end{abstract}

\maketitle


The intense interest in iridium oxides, or iridates, arises from a number of competing interactions of similar magnitude  \cite{Okamoto2007, BJKIM2008, BJKim2009, Jackeli2009, Shitade2009, Pesin2010, Jiri2010, Sr2IrO4-RIXS, Clancy2012}. While the on-site Coulomb interaction is the dominant energy scale in 3$d$ transition metal oxides, the spin-orbit coupling (SOC) is largely ignored. On the other hand, for 5$d$ elements such as Ir, the SOC becomes significant, and in fact plays a dominant role. A good example is \SIO, whose electronic states are well described by \jeff=1/2 states arising from the spin-orbit split $t_{2g}$ levels \cite{BJKIM2008,BJKim2009,Sr2IrO4-RIXS}.

One of the most intensely scrutinized families of iridates is the honeycomb lattice family \AIO\ ($A$=Na,Li) \cite{Shitade2009,Jiri2010,Yogesh2010,Yogesh2011,Reuther2011,Khomskii2012}. Originally thought of as topological insulator \cite{Shitade2009}, these materials are now believed to be Mott insulators \cite{Yogesh2010,Yogesh2011}. A recent calculation though suggests that uniaxial strain might still drive the system to topological insulating behavior \cite{Choong}. Furthermore, these materials could be described with the Kitaev-Heisenberg model \cite{Jiri2010,Yogesh2011}, in which bond-dependent Kitaev interaction are realized and support various types of topological phases. The applicability of such intriguing theoretical possibilities to real system crucially depends on the \jeff\ physics arising from strong SOC. However, the experimental situation seems to be far from clear. In particular, structural refinements find a sizable trigonal distortion of the IrO$_6$ octahedra \cite{Coldea2012,GangCao2012}, which will produce crystal field splittings within the $t_{2g}$ manifold. If the splitting is comparable to the SOC, the \jeff=1/2 states will mix with \jeff=3/2 states and the relevant microscopic model becomes quite different from the ideal \jeff\ physics \cite{Subhro2011,Khomskii2012}, preventing the Kitaev-Heisenberg model from being realized \cite{Jiri2010,Reuther2011,Yogesh2011}. Recent theoretical studies have even suggested that the ground state has a large contribution from the \jeff=3/2 state \cite{Lovesey2012}.

Therefore, it is of great importance to elucidate the underlying electronic structure of \NIO\ experimentally. In particular, the spectroscopic investigation of excitations between spin-orbit split \jeff\ states can provide us with direct information regarding the size of the crystal field splitting with respect to the typical SOC energy scale in iridates (0.4-0.5~eV). In the case of \SIO, such excitations between \jeff=3/2 to \jeff=1/2 were observed around 0.6-0.8~eV in the resonant inelastic x-ray scattering (RIXS) data \cite{Sr2IrO4-RIXS}, which is accounted for in the quantum chemical calculation by Katukuri et al. \cite{Vamshi}.  The splitting within these ``spin-orbit" excitations arises due to non-zero tetragonal crystal fields, and is much smaller ($\sim 0.1$~eV) than the SOC, justifying the \jeff\ description of Sr$_2$IrO$_4$.

In this Letter, we present a comprehensive picture of the low energy
electronic structure of $\rm Na_2IrO_3$ and $\rm Li_2IrO_3$, based on Ir
$L_3$-edge RIXS experiments. Our high-resolution RIXS measurements allow us to resolve the crystal field splitting of the \jeff=3/2 states due to the trigonal distortion, which is determined to be about 110 meV in both compounds. This energy scale agrees very well with quantum chemical calculations, and is much smaller than the typical value for SOC, validating the \jeff\ picture in these compounds. We have also studied momentum dependence of the insulating gap; the observed flat dispersion of the insulating gap is consistent with what is expected from a significant Coulomb interaction in both compounds. Taken together, we argue that just as \SIO, the honeycomb \AIO\ iridates can be described as spin-orbit Mott insulators \cite{BJKIM2008,Subhro2011,Choong,Comin}.

\begin{figure}[htb]
\includegraphics[width=\columnwidth]{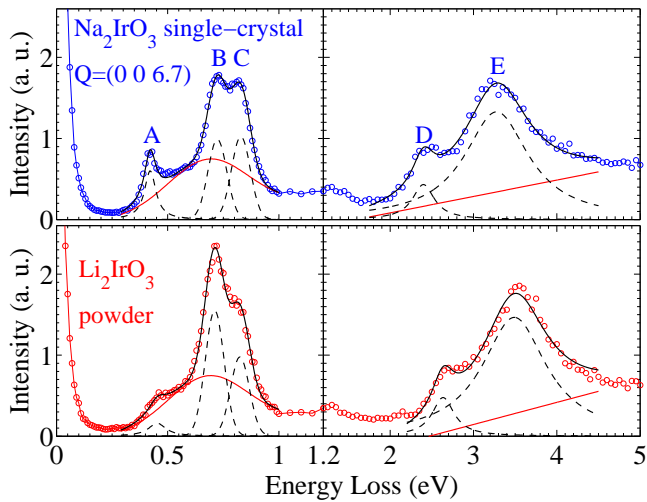}
\caption{\label{fig01}(Color online) Top: Wide energy range RIXS
spectrum for a single-crystal sample of \NIO\ at {\bf Q} = (0 0 6.7) obtained with $E_i$ = 11.217
keV. Note the different scale used for left and right panel. Bottom: RIXS spectrum for \LIO\ powder sample at $\rm |{\bf Q}| \approx 8$ $\rm \AA^{-1}$, obtained with the same $E_i$. All spectra
were measured at room temperature. The black dashed curves are the result of a fit (see text) and the red solid lines represent the background. }
\end{figure}


The RIXS experiment was carried out at the
Advanced Photon Source using the 30ID MERIX  and 9ID RIXS spectrometer. A spherical (1~m radius) diced Si(844) analyzer and Si(844) secondary monochromator were used to obtain overall energy resolution (FWHM) of $\sim 35$~meV. To minimize the elastic background, most of
the measurements were carried out in a horizontal scattering geometry
near {\bf Q} = (0 0 6.7), for
which the scattering angle 2$\theta$ was close to 90$^\circ$. We use the $C2/m$ notation for the lattice \cite{Coldea2012,GangCao2012}. A
single crystal of Na$_2$IrO$_3$ and a polycrystalline samples of \LIO were grown by the solid state synthesis method, previously described in detail \cite{Yogesh2010,Yogesh2011}. The \NIO\ crystal was plate-like with a flat shiny surface, the surface normal was in the (001) direction.


The RIXS process at the $L_3$-edge of Ir (or any other $d$ electron
system) is a second order process consisting of two dipole
transitions ($2p \rightarrow 5d$ followed by $5d \rightarrow 2p$).
Therefore, it is especially valuable for detecting excitations
between the $d$-levels and has been extensively
utilized in the study of $3d$ transition metal compounds \cite{Moretti2011,Ghiringhelli,Vernay,Ghiringhelli2004,Luuk}. Recent
instrumental advances have made it possible to measure
collective magnetic excitations \cite{Braicovich09,Sr2IrO4-RIXS}.
In \AIO, Ir$^{4+}$ ions are in the $5d^5$ configuration in a
slightly distorted octahedral environment of oxygen ions, with the
edge-sharing IrO$_6$ octahedra forming a honeycomb net. Due to the
octahedral crystal field, there exists a fairly large splitting
($10Dq$) between the $t_{2g}$ and $e_g$ states. Since the $5d$
orbitals are spatially more extended than the $3d$ orbitals, the
$10Dq$ value is expected to be much larger. Indeed, in our RIXS
investigations of various iridium compounds, well separated $t_{2g}$
and $e_g$ states have been observed, with the $10Dq$ value typically
about 3 eV \cite{CuIr2S4}.

In Fig. \ref{fig01}, a representative high-resolution RIXS spectrum of \NIO\ is plotted on a wide energy scale. This scan was
obtained at room temperature and plotted as a function of energy loss $(\hbar\omega=E_i-E_f)$. The incident energy, $E_i=11.217$ keV, was
chosen to maximize the resonant enhancement of the spectral features
of interest below 1~eV. A broad and strong feature is observed at $2$-$4$ eV  and other sharper features are observed below 1 eV, corresponding to
$d$--$d$ transitions from occupied $t_{2g}$ states into the empty $e_g$ and $t_{2g}$ levels, respectively. Also plotted in the figure is the room temperature data of polycrystalline \LIO. Lack of significant momentum dependence of these $d$--$d$ excitations (shown later in Fig.~3) allows one to directly compare the peak positions between the single crystal and powder samples. The spectra were fit to 5 peaks (labeled A-E), as shown by the black dashed lines. The low energy excitations can be fit to three peaks, two Gaussians (B and C) of the same width and one Lorentzian (A) on top of a broad background (Gaussian). Two Lorentzian functions with sloping background were used to fit the  higher energy excitations (D and E). The resulting peak positions are listed in Table \ref{tab1}.

\begin{table}
\caption{\label{tab1}
RIXS and MRCI+SOC excitation energies ($C2/m$ structure) for 213 iridates (eV).
}
\begin{ruledtabular}
\begin{tabular}{lllll}
        &Na213    &Na213      &Li213    &Li213     \\
        &RIXS     &MRCI       &RIXS     &MRCI      \\
\hline
Peak A  &0.42(1)  &--         &0.45(2)  &--        \\
Peak B  &0.72(2)  &0.82       &0.72(2)  &0.80 \\
Peak C  &0.83(2)  &0.89       &0.83(2)  &0.97 \\
Peak D  &2.4(1)   &2.8--3.4   &2.6(1)   &3.1--3.7\\
Peak E  &3.3(1)   &3.8--4.1   &3.5(1)   &4.1--5.0\\
\end{tabular}
\end{ruledtabular}
\label{tab:tab1}
\end{table}

To clarify the nature of the excitations revealed by RIXS, we have carried out
multiconfiguration self-consistent-field and multireference configuration-interaction (MRCI)
calculations \cite{book_QC_00} on clusters consisting of one central IrO$_6$ octahedron, all adjacent Na or Li ions, and the three nearest-neighbor (NN) IrO$_6$ octahedra (see Ref. \citenum{Vamshi} and Supplemental Material for details).
Local $d$--$d$ transitions are computed only for the central IrO$_6$ octahedron while the NN octahedra are explicitly included in the cluster for providing an accurate description of the nearby charge distribution. Two different lattice configurations are considered, i.e., the $C2/c$ structure \cite{Yogesh2010,Kobayashi2003} and also the $C2/m$ arrangement proposed more recently \cite{Coldea2012,GangCao2012,Malley2008}.

Results of spin-orbit MRCI (MRCI+SOC) calculations using the $C2/m$ configuration \cite{Coldea2012}
are listed for Na$_2$IrO$_3$ in the third column of Table I.
The MRCI+SOC data fit the experiment reasonably well, with peaks B and C corresponding to \jeff=3/2
to \jeff=1/2 electronic transitions.
Above 2.5 eV, the MRCI+SOC results indicate multiple $t_{2g}$ to $e_g$ excitations
displaying a two-peak structure reminiscent of the D and E features in the RIXS spectra.
However, MRCI+SOC seems to overestimate somewhat the relative energies of those latter features.
Interestingly, for the alternative $C2/c$ structure of Na$_2$IrO$_3$ \cite{Yogesh2010},
the splitting between the two doublets originating from the  \jeff=3/2 quartet in an ideal
octahedral environment is much larger and the position of the C peak is overestimated by 0.25 eV in
the MRCI+SOC treatment.
Since the deviations from the experimental data are in this case larger, the MRCI+SOC results for
$C2/c$ symmetry are not listed in Table I. The $t_{2g}$ splittings in calculations with no SOC are in fact as large
as 0.6 eV for the $C2/c$ structure of Na$_2$IrO$_3$, which gives rise to a highly uneven
admixture of $t_{2g}$ components in the spin-orbit calculations.
In contrast, for the $C2/m$ configuration, the $t_{2g}$ splittings are about 0.1 eV and the
three different $t_{2g}$ hole configurations contribute with similar weight to the spin-orbit
ground-state wave function (see Table II).

For Li$_2$IrO$_3$, the calculations correctly reproduce the shift to higher energies of the $t_{2g}$
to $e_g$ transitions relative to those in \NIO. The discrepancy between the experimental values and the MRCI+SOC results (e.g., peak C) could be caused by the uncertainty in the structural model used for this calculation ($C2/m$ from Ref.~\cite{Malley2008}). Since local structural disorder is not easily captured in the regular diffraction data, local structure probes such as pair-distribution function (PDF) measurements can sometimes be useful for clarifying the structural details. We have carried out X-ray PDF studies on \LIO\ and \NIO\ powder samples. Details of these measurements and the comparison of the two structures are reported in the Supplemental Material. Except for the overall lattice contraction, the \LIO\ PDF seems to be well described by the $C2/m$ symmetry, eliminating the local structural disorder as a possible explanation. Most likely cause of the structural uncertainty is the oxygen position, since x-ray structural probes are not particularly sensitive  to light elements like oxygen \cite{Kobayashi2003,Malley2008}. We note that the latest refinements using both powder neutron and single crystal x-ray data on \NIO\ do show important differences compared to earlier x-ray powder diffraction data and the MRCI+SOC results are very different for the two structures. Better structural refinements using neutron diffraction would reduce the oxygen position uncertainty in \LIO\ and could improve the agreement between our MRCI+SOC calculation and the experiment.

One of our main findings is that the splitting of the strong RIXS peak located at 0.7-0.8 eV is due to the trigonal distortion which is well corroborated with our MRCI+SOC calculations. The fact that this splitting (110 meV) is much smaller than a SOC of 0.4-0.5 eV strongly supports that these excitations are transitions from crystal-field-split \jeff=3/2 levels to the \jeff=1/2 state (labeled spin-orbit exciton in Ref.~\cite{Sr2IrO4-RIXS}).
Given that the optical gap in this material is about 350~meV \cite{Comin} and that there is no such excitation in the MRCI+SOC calculations which only look at  on-site $d$--$d$ excitations, it is reasonable to associate feature A at low energy as arising from the excitation of a particle and hole pair across the charge gap. Additional periodic density functional theory (DFT) calculations shows that a moderate size $U$ and SOC can indeed open a (Mott) gap of 300-400 meV, in accordance with the experimental observation (see Supplemental Material).

\begin{table}
\caption{Percentage contributions of the different Ir $5d^5$ configurations to the lowest on-site $d$--$d$ excited states in \NIO, as obtained from MRCI+SOC calculations.}
\begin{ruledtabular}
\begin{tabular}{llll}
Energy (eV) & 0 & 0.82 & 0.89 \\
\hline
$d_{xy}^2d_{yz}^2d_{zx}^1$ & 38.7 & 24.3 & 32.2 \\
$d_{xy}^2d_{yz}^1d_{zx}^2$ & 34.7 & 60.3 & 24.7 \\
$d_{xy}^1d_{yz}^2d_{zx}^2$ & 26.6 & 15.4 & 43.1 \\
\end{tabular}
\end{ruledtabular}
\label{tab:tab2}
\end{table}

\begin{figure}[htb]
\includegraphics[width=\columnwidth]{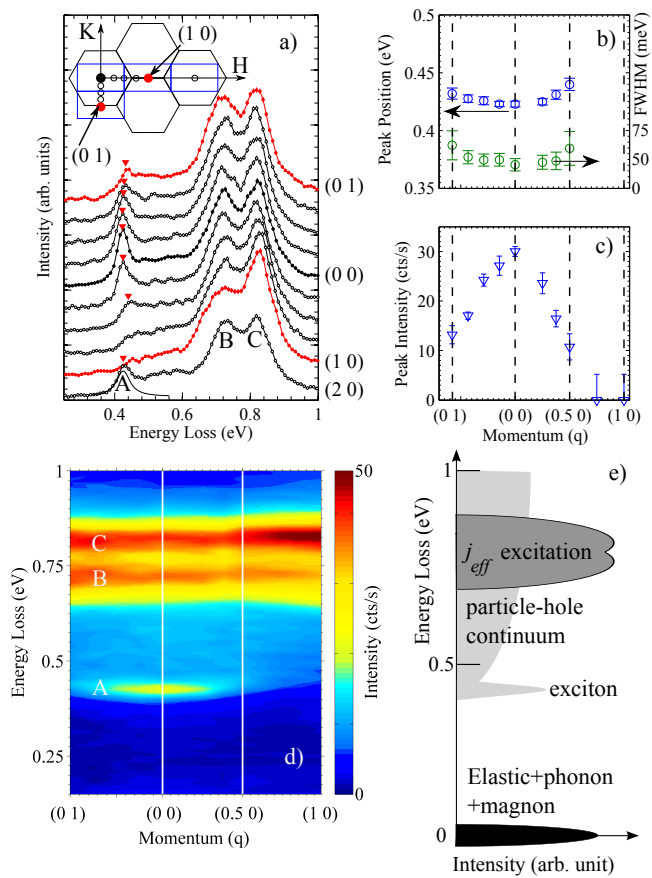}
\caption{(Color online) (a) Momentum dependence of the low energy RIXS spectra of \NIO\ obtained at $T=9$~K. The inset shows a schematic diagram of the (H K 0) reciprocal space plane. The Brillouin zones (BZ) corresponding to the monoclinic unit cell are blue rectangles. For comparison, we also plot the BZ of the honeycomb net in black. The circles are the points where RIXS spectra are taken. The low energy peaks denoted with red triangles are fit to a Lorentzian, and the momentum dependence of (b) the peak position and width, and (c) the peak intensity are shown. (d) Same data are plotted in false color scale. (e) Schematics of electronic excitations in \AIO\ determined from our RIXS measurements.}\label{fig03}
\end{figure}

The nature of the charge excitation gap can be further revealed by its momentum dependence. In Fig.~\ref{fig03}, we plot the momentum dependence of the low energy peaks (A-C) in \NIO. In the honeycomb plane, the magnetic ordering doubles the unit cell \cite{Xuerong2011}, and correspondingly the first Brillouin zone (BZ) becomes smaller. Two different BZ schemes are illustrated in the inset of Fig.~\ref{fig03} (a) to aid the comparison. We will use the rectangular BZ notation. Note that the two high symmetry directions of interest, the {\bf q}=(H~0) and {\bf q}=(0~K) in rectangular notation, correspond to the $\Gamma-M$ and $\Gamma-K$ directions in the honeycomb plane, respectively. One can see that the overall momentum dependence of the peak positions is very small, except for peak A. To investigate the behavior of peak A in detail, the low energy portion of the spectra was fit to a Lorentzian peak. Since the peak seems to disappear at {\bf q}=(1~0), we have used the spectrum at this {\bf q} as an empirical background. The fitting results for peak positions, widths, and intensities are shown in Fig.~\ref{fig03}(b)-(c). The width and peak position remains almost unchanged ($\approx 10$~meV dispersion), but the intensity is strongly peaked around the BZ center. This can be clearly seen in the pseudocolor plot of the spectra shown in Fig.~\ref{fig03}(d), in which a strong peak around {\bf q}=(0 0) and 0.42 eV is contrasted with the {\bf q}-independent features B+C. In addition, one can see that the spectral weight changes abruptly around 0.4~eV, confirming that this is the particle-hole continuum boundary. Based on our RIXS results, the electronic excitations in \AIO\ can be summarized as shown in Fig.~\ref{fig03}(e).

It is clear from this observation that the insulating gap is direct (minimum gap at $\Gamma$). The relatively flat dispersion observed in our data is also consistent with the DFT calculation which suggests that the correlation effect makes the bandwidth smaller, leading to an almost dispersionless charge gap. The sharpness in energy and momentum of peak A is quite reminiscent of the excitonic behavior of the BZ center particle-hole excitation across the charge-transfer gap in the insulating cuprate $\rm La_2CuO_4$ \cite{Ellis2008}. This suggests that an extra nearest-neighbor Coulomb interaction $V$ (in addition to the on-site interaction $U$) might be important for modelling this material. Sizable $V$ could promote the tendency towards exciton binding and also further narrow the bandwidths. The smaller intensity of the charge gap feature in \LIO\ compared to \NIO\ could be due to the fact that the \LIO\ data are powder averaged. However, one cannot rule out the possibility of weaker $V$ in \LIO\ as compared to \NIO.

Another interesting aspect of our data is that the dispersion of the gap appears to follow the underlying honeycomb lattice rather than the crystallographic/magnetic unit cell. This is clearly observed by the spectrum obtained at {\bf q}=(2 0). While (2 0) is the next BZ center along the $\Gamma-K$ direction, (1 0) is on the zone boundary; peak A disappears at (1 0) but recovers its intensity at the {\bf q}=(2 0) position. Additional momentum dependence data, reported in the Supplemental Material, shows the lack of momentum dependence along the $L$-direction (perpendicular to the honeycomb plane). Therefore, the electronic structure of \NIO\ seems to be quite well described as that of a 2D honeycomb lattice.

It is worth comparing the observed low energy RIXS spectrum with that of \SIO. In \SIO, a low energy magnon was observed below 200 meV, while highly dispersive excitations were observed between 0.4 eV and 0.8 eV. This latter band of excitations is composed of particle-hole excitation across the Mott gap and spin-orbit excitations from \jeff=3/2 states to the \jeff=1/2 states. Because of the smaller single-particle band width in \AIO\ (see DFT calculations in Ref. \citenum{Choong})
, the ``\jeff\ excitation" in \NIO\ is almost dispersionless, unlike the highly dispersive counterpart in \SIO. Perhaps an even more significant difference is the well separated energy scale of the \jeff\ excitation and the particle-hole continuum in \NIO. These two energy scales are very similar in \SIO, but the large separation in \NIO\ allows one to investigate these two types of excitations separately.


To summarize, we have carried out a resonant inelastic x-ray scattering
investigations of electronic excitations in \NIO\ and \LIO.
We observe three well-defined features below 1 eV and a broad two peak
feature at 2-5 eV. By comparing our observation with quantum chemical and density
functional theory calculations, we associate these features with
$d$--$d$ transitions. Specifically, the high energy excitations are from $t_{2g}$ to $e_g$ excitations, while the low energy excitations around 0.7-0.8 eV are excitations from \jeff=3/2  to \jeff=1/2 states. The splitting of the latter feature arising from the trigonal crystal field is about 110~meV, much smaller than the spin-orbit coupling energy scale of Ir compounds, which validates the applicability of \jeff\ physics in \AIO. In addition, we observe a lower energy excitation around 0.4 eV, which shows very little momentum dependence and is associated with the particle-hole excitation across the Mott gap; the ``excitonic" behavior of this peak suggests the nearest-neighbor Coulomb interaction $V$ is sizable. We conclude that the electronic structures of both \NIO\ and \LIO\ are similar and these systems can be described as spin-orbit Mott insulators.

We would like to thank Y. B. Kim and S. Bhattacharjee for fruitful discussions and Doug Robinson for technical assistant during the PDF measurements.
Research at the U. of Toronto was supported by the NSERC, CFI, and OMRI.  This research benefited from the RIXS collaboration supported by the Computational Materials and Chemical Sciences Network (CMCSN) program of the Division of Materials Science and Engineering, U.S. Department of Energy, Grant No. DE- SC0007091. Use of the APS was supported by the U. S. DOE,
Office of Science, Office of BES, under Contract No. W-31-109-ENG-38. Work performed at Brookhaven National Laboratory was supported by DOE, Office of Science, Division of Materials Science under contract No. DE-AC02-98CH10886.
Y.-J.K. was supported by the KOFST through the Brainpool program. H-S Kim and J. Yu were supported by the NRF through the ARP (R17-2008-033-01000-0). H-S Kim would like to acknowledge the support from KISTI supercomputing center through the strategic support program for the supercomputing application research (No. KSC-2010-S00-0005). S. Manni acknowledges support from the Erasmus Mundus Eurindia Project.

\section{Supplemental Material}
\subsection{S1. First principle electronic structure calculations}

In order to understand the origin of the low energy excitations around 0.4 eV,
we have carried out periodic density functional theory (DFT) electronic-structure calculations. We have used the DFT code OpenMX \cite{S_openmx} based on the linear
combination of pseudo-atomic orbital(LCPAO) formalism \cite{S_ozaki},
 the Perdew-Burke-Ernzerhof GGA-functional,  $8\times 6\times
8$ k-points grids within the Brillouin zone, and 300 Ry for the real-space
grid. The  SOC is treated via a fully-relativistic $j$-dependent pseudopotential in the
non-collinear DFT formalism \cite{S_macdonald,S_bachelet,S_theurich}. We applied  the $C2/m$ monoclinic crystal structure reported in Ref.~\cite{S_Coldea2012} and the zigzag-type magnetic order suggested in Ref.~\cite{S_Xuerong2011} for \NIO. For \LIO, we used the same magnetic stucture as for \NIO\ and structural data from Refs.~\cite{S_Malley2008}.

Densities of states (DOS) from GGA and GGA+U  \cite{S_han} calculations are
plotted in Fig.~\ref{fig02}. The solid line is the total DOS and the filled shaded area represents partial $d$-orbitals DOS. From top to bottom, the calculations
were done with no SOC or $U$, with SOC, and with both SOC and $U$.
As expected the $5d$ states are dominant near the Fermi level, although there exists some modest amount of hybridization between Ir $5d$ and O $2p$ states. A moderate size $U$ can open a (Mott) gap of 300-400 meV, in accordance with the experimental observation. Compared to \NIO, the bandwidths in \LIO\ are a little bit larger but the nature of the gap is essentially the same.

\renewcommand{\thefigure}{S1}
\begin{figure}[htb]
\includegraphics[width=\columnwidth]{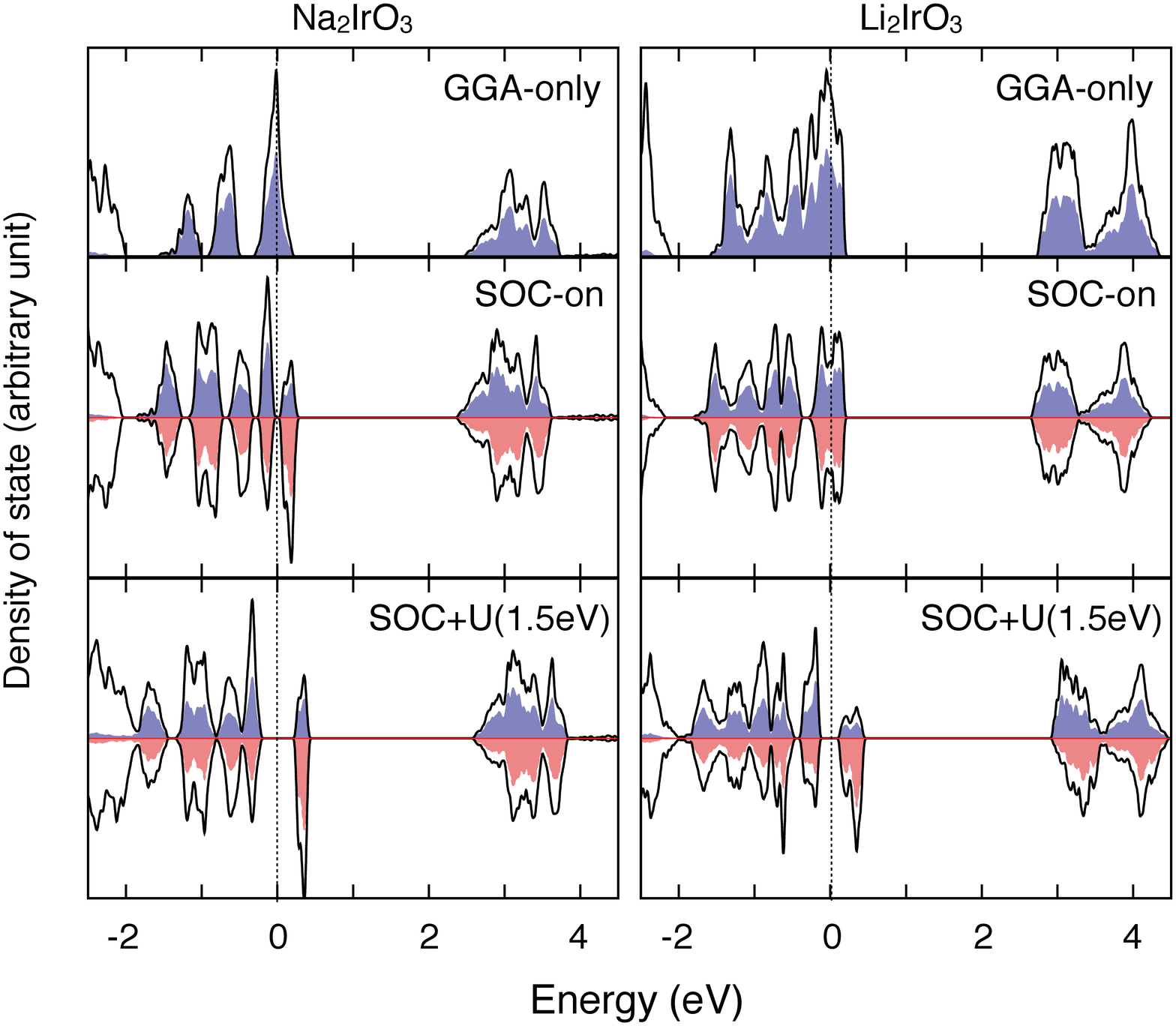}
\caption{(color online) The evolution of the DOS for both \NIO\ and \LIO\ upon inclusion of spin-orbit coupling and on-site correlation.  The solid line is the total DOS and the filled shaded area represents partial $d$-orbital DOS.}\label{fig02}
\end{figure}

\subsection{S2. Quantum chemical calculations}
To investigate in detail the electronic structure and the essential interactions in the \AIO\ iridates, we further performed correlated quantum chemical calculations. In the quantum-chemical study, we employed multiconfiguration self-consistent-field and multireference configuration-interaction (MRCI) methods \cite{S_book_QC_00} as implemented in the {\sc molpro} package \cite{S_molpro_brief}. The calculations were performed on fragments consisting of one central IrO$_6$ octahedron for which the local $d$--$d$ transitions are explicitly computed plus all nearest-neighbor (NN) octahedra, three in \AIO, and adjacent Na or Li ions. To simplify the analysis of the wave functions, the NN Ir$^{4+}$ ions were modeled as closed-shell Pt$^{4+}$ species \cite{S_Nikolai,S_Vamshi}. The remaining part of the crystal is represented as an array of point charges that reproduce the Madelung field in the cluster region. Effective core potentials and basis sets as described in earlier investigations
on Sr$_2$IrO$_4$, Ba$_2$IrO$_4$, and CaIrO$_3$ were used \cite{S_Nikolai,S_Vamshi}.

\subsection{S3. Additional momentum dependence}

In many layered systems, such as the cuprates \cite{S_Kenji2005}, momentum dependence along the $L$-direction is expected to be small. However, recent DFT calculations have shown dispersion of the optical gap along the $L$-direction \cite{S_Comin2012}. In order to investigate this we measured the momentum dependence along the {\bf Q} = (0 0 L) direction
in \NIO\ at T = 9 K. These spectra were taken with the same high-resolution setup as the one in Fig. 3 (a). The RIXS spectra in Fig. \ref{Ldep} show no observable dispersion, supporting the 2D nature of \NIO.
We have also measured the momentum dependence of the high energy excitation (labelled D and E in Fig. 1). These spectra were taken with an overall resolution of $\sim\!\!150$ meV. Fig. \ref{eg_Qdep} show RIXS spectra taken along the {\bf Q} = (-H -H 6.9) direction in \NIO\ at T = 9 K. No observable changes were seen.

\renewcommand{\thefigure}{S2}
\begin{figure}[htb]
\includegraphics[width=\columnwidth]{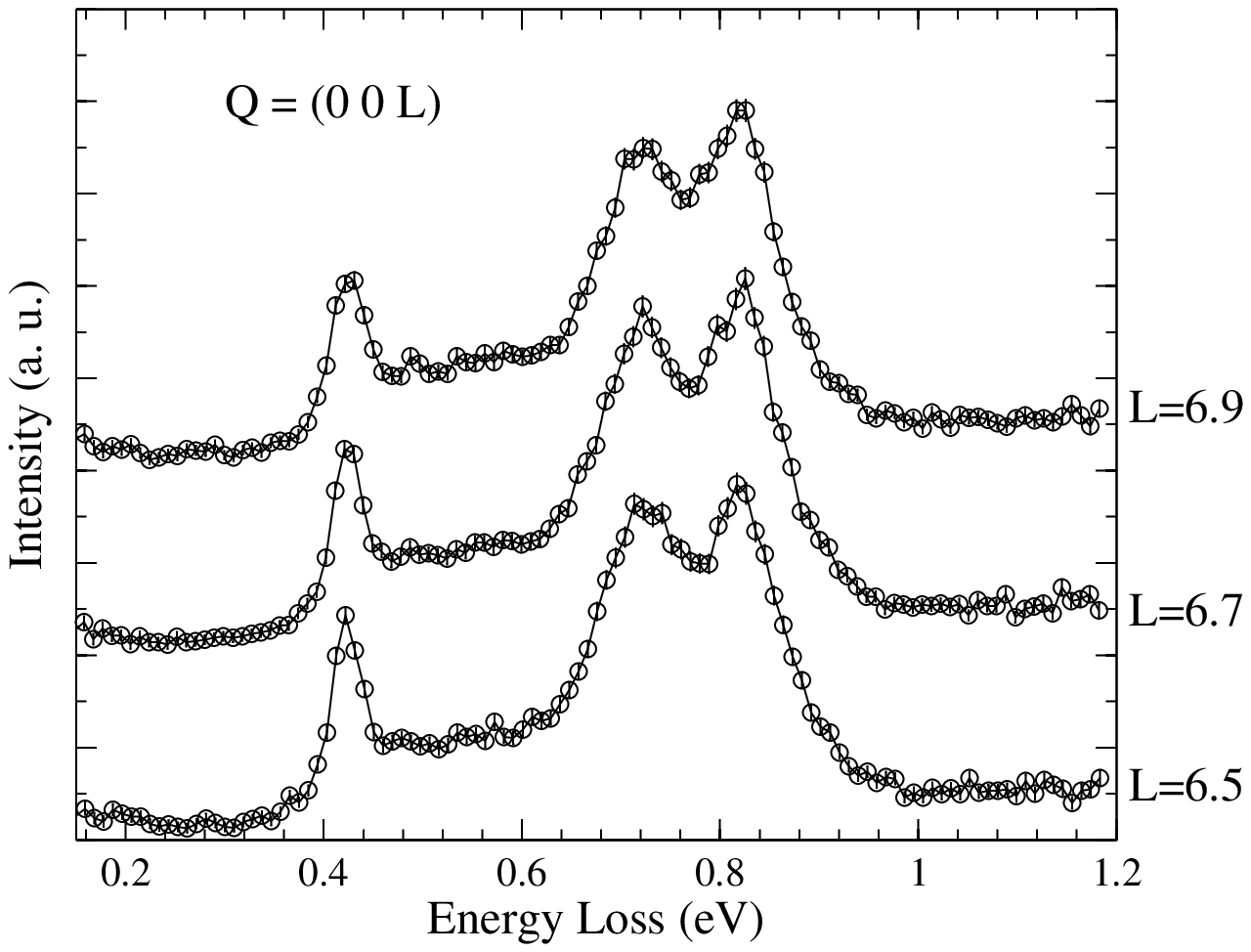}
\caption{ Momentum dependence of the low energy RIXS spectra in \NIO\  along the  {\bf Q} = (0 0 L) direction. All data sets collected at $T=9$~K.}\label{Ldep}
\end{figure}

\renewcommand{\thefigure}{S3}
\begin{figure}[htb]
\includegraphics[width=\columnwidth]{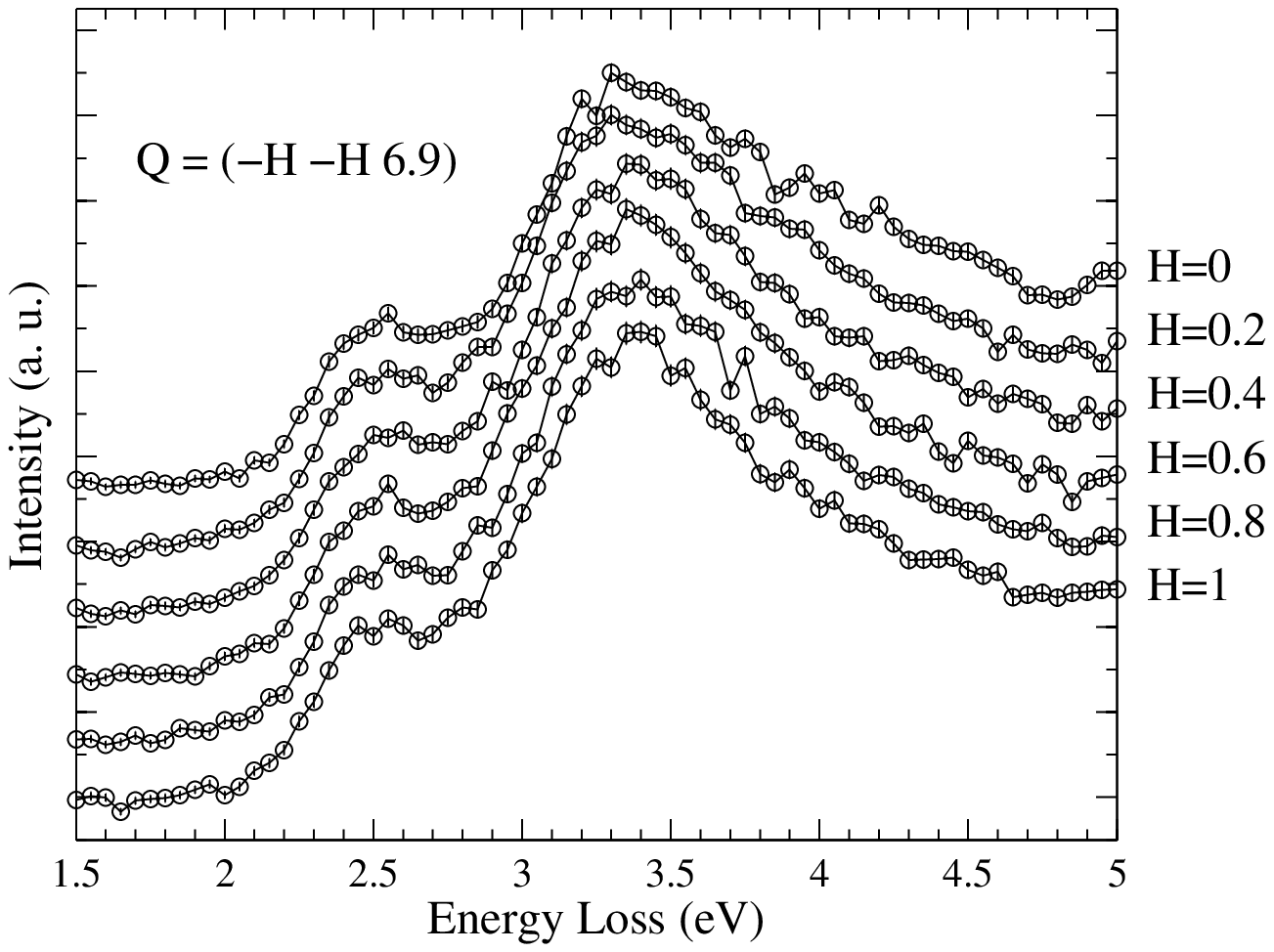}
\caption{ Momentum dependence of the high energy RIXS spectra in \NIO\ along the {\bf Q} = (H H 6.9) direction. All data sets collected at $T=9$~K with a resolution of  $\sim\!\!150$ meV (FWHM). }\label{eg_Qdep}
\end{figure}

\renewcommand{\thefigure}{S4}
\begin{figure}[htb]
\includegraphics[width=0.7\columnwidth,angle=-90]{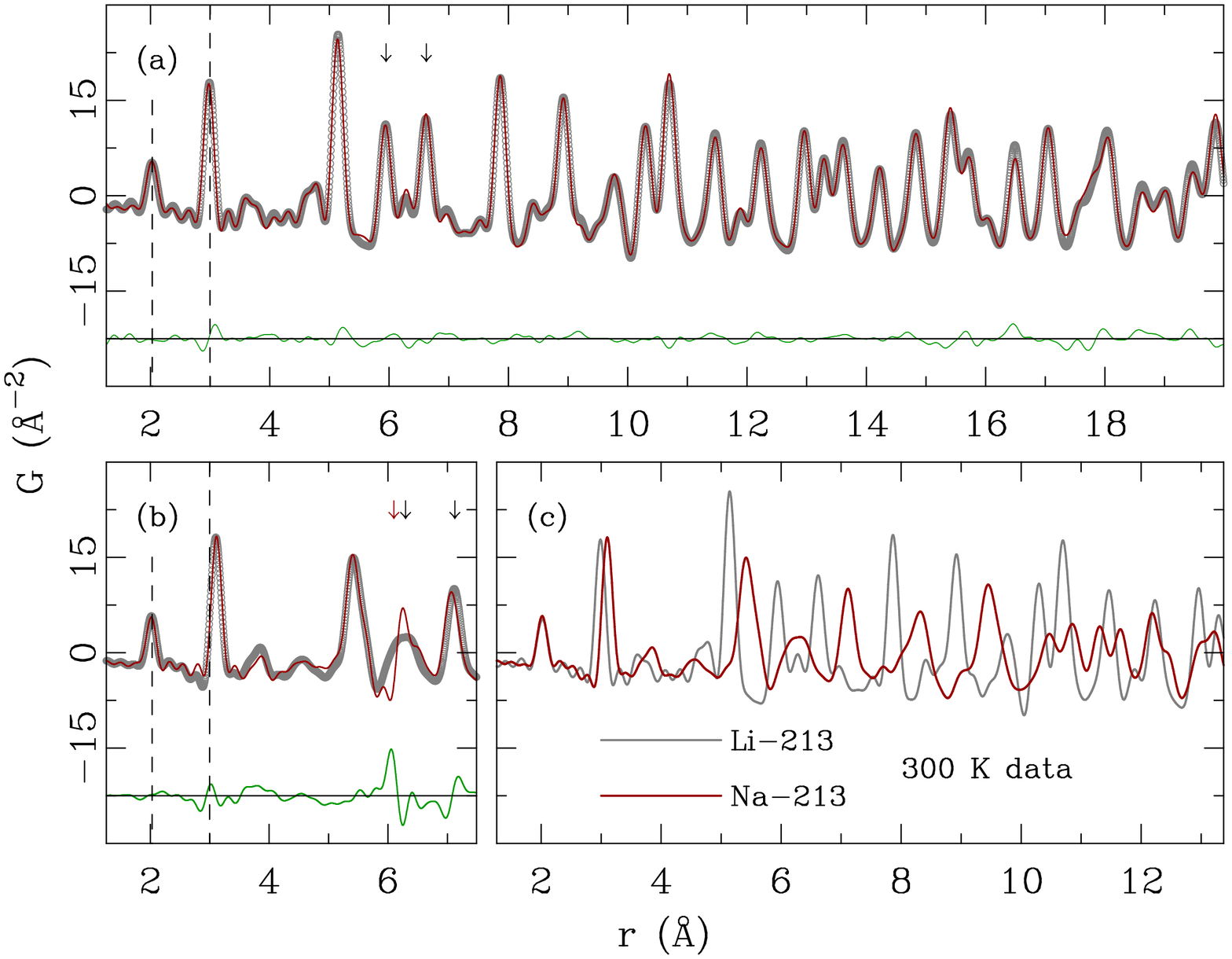}
\caption{ (color online) Atomic PDF of \LIO\ and \NIO\ at 300 K. (a) \LIO\ data (gray open symbols) and refined $C2/m$ model (red solid line), with the difference curve (green solid line) that is offset for clarity. (b) The same as (a) but for \NIO. (c) Direct comparison of \LIO\ and \NIO\ experimental PDF data at 300 K, scaled such that the intensities of the first PDF peaks overlap. Dashed vertical lines in (a) and (b) correspond to the first two sharp PDF peaks in \LIO. Arrows denote specific PDF features discussed in the text.   }\label{PDF}
\end{figure}

\subsection{S4. X-Ray Atomic Pair Distribution Function (PDF) Measurements}

Atomic PDF measurements were performed at 300 K at the 6-ID-D beamline of the Advanced Photon Source at Argonne National Laboratory, utilizing a General Electric amorphous silicon image plate (IP) detector. A monochromatic incident x-ray beam $0.5\rm{mm}\times 0.5\rm{mm}$ in size was used, conditioned to have energy of 74.353 keV ($\lambda$=0.1668 $\rm{\AA}$). Finely pulverized Li$_2$IrO$_3$ and Na$_2$IrO$_3$ samples were packed in cylindrical polyimide capillaries 1.0 mm in diameter and sealed at both ends. The IP detector was mounted orthogonally to the beam path with a sample to detector distance of 221.87 mm, as calibrated using a ceria (CeO$_2$) standard sample \cite{S_Billinge2003}. The 2D diffraction data were integrated and converted to intensity versus $2 \theta$ using the software FIT2D \cite{S_Hauserman1996}, where $2 \theta$ is the angle between the incident and scattered x-ray beam. The intensity data were corrected and normalized \cite{S_Billinge2003_Book} using the program PDFgetX2 \cite{S_Billinge2004} to obtain the total scattering structure function, $F(Q)$ and its Sine Fourier transform, i.e. the atomic PDF, $G(r)$. $Q_{\rm{max}}$ of 25 $\rm{\AA}^{-1}$ was used in the transform. The PDF analysis was carried out using the program PDFgui \cite{S_Billinge2007}. Atomic PDF yields a histogram of interatomic distances in a material, and provides structural information on short, intermediate, and long range lengthscales \cite{S_Billinge2003_Book}.

The PDF data of \LIO\ could be explained over a broad $r$-range within the $C2/m$ model with all crystallographic sites fully occupied, with no indication of local structural distortions being present, as evident in Fig. \ref{PDF} (a).  Structural parameters, as refined over a 1.25-20.0 $\rm{\AA}$ range, can be summarized as follows: a=5.172(1) $\rm{\AA}$, b=8.926(2) $\rm{\AA}$, c=5.122(2) $\rm{\AA}$, $\beta=109.91(4)^{\circ}$, with Ir at 4g (0.5, 0.167(1), 0), Li1 at 2a (0,0,0), Li2 at 2d (0.5, 0, 0.5), Li3 at 4h (0.5, 0.31(6), 0.5), O1 at 8j (0.752(5), 0.173(2), 0.769(4)), and O2 at 4i (0.707(8), 0, 0.260(7)) atomic positions. The Ir-O distances and Ir-O-Ir bond angles range from 2.01 to 2.04 $\rm{\AA}$ and 93.6 to 94.8$^{\circ}$ respectively, while Ir-Ir nearest neighbor distances are in the range from 2.984 to 2.992 $\rm{\AA}$.

On the other hand, while the fit of the $C2/m$ crystallographic model to the PDF data of \NIO\ is consistent with the low-$r$ PDF features, significant discrepancies are observed beginning at around 6 $\rm{\AA}$, as can be seen in Fig. S4 (b), and no convergence could be achieved in broad $r$-range fits irrespective of refinement strategies attempted. The $C2/m$ model is inadequate in describing the intermediate structure of \NIO\, and the actual symmetry of the intermediate range structure is lower. This PDF is consistent with there being an appreciable amount of disorder in the \NIO\ sample compared to \LIO. To qualitatively illustrate this, it is useful to directly compare the experimental PDFs of the two \AIO\ systems, Fig. \ref{PDF} (c). Local structural features are compared first. The first sharp PDF feature containing nearest neighbor Ir-O distance in \NIO\ occurs at around 2 $\rm{\AA}$, and coincides with that of \LIO. The next sharp PDF feature, which includes the Ir-Ir near neighbor peak, in \NIO\ occurs at an observably larger distance in \NIO\ (3.105 $\rm{\AA}$) than in \LIO\ (2.985 $\rm{\AA}$), indicating that the Ir-rings of the honeycomb network are appreciably larger, and reflecting the lattice expansion on going from smaller Li to larger Na. Despite the shift in peak positions, the relative intensities of these short range PDF features are comparable for the two samples, and have comparable peak widths, suggesting that the underlying local environments are very similar in the two systems.

We now consider the intermediate lengthscale in PDFs shown in Fig. S4 (c). What is immediately apparent is that starting from about 5-5.5 $\rm{\AA}$ the PDF features of \NIO\ are visibly broader than those in \LIO\, indicative of disorder in the former. Notably, while the $C2/m$ model explains the features in the PDF of \LIO\ in 6-7 $\rm{\AA}$ range (marked by black arrows in Fig. \ref{PDF} (a)), the same model fails to explain the corresponding features in the \NIO\ data. The two features marked by arrows in \LIO\ PDF are both sharp and of approximately equal intensity. In contrast, the corresponding PDF peaks in \NIO\ PDF (marked by black arrows in Fig. \ref{PDF} (b)) have very different intensities. The PDF feature at around 6 $\rm{\AA}$ in \NIO\ is rather broad, with extra intensity appearing at the low-$r$ side of the peak (indicated by red arrow in Fig. \ref{PDF} (b)), and is clearly multi-component. This is precisely the region where the broad $r$-range fits first fail to explain the PDF of \NIO, as these features are not incorporated into the $C2/m$ structural model. Although extracting  the details of the actual intermediate structure of \NIO\ from the present 300 K data is a difficult task, further hampered by the thermal broadening that lowers the PDF resolution and requiring elaborate analysis of the low-temperature PDF data for full characterization, it is tempting to speculate about the origin of the observed features. Considerations of the crystallographic model reveal that in the 6 $\rm{\AA}$ range there are contributions from the third Ir-Ir neighbors in the honeycomb, as well as contributions involving Ir-Ir pairs belonging to two successive honeycomb sheets. It is plausible that the observed discrepancies originate from stacking faults suggested in reference \cite{S_Coldea2012}, where appreciable rods of diffuse scattering were observed in the single crystal scattering experiments. However, if we assume that stacking faults occur randomly at the $<10\%$ level, a sharp signal in PDF such as that seen in the difference curve shown in Fig. \ref{PDF} (b) is difficult to explain. Considering that the PDF feature around 5.3~$\rm{\AA}$ in \NIO, containing the next near neighbor Ir-Ir distance of the honeycomb, is already broad compared to its \LIO\ counterpart, we cannot rule out the possibility that in addition to the distortions generated by stacking faults there are also in-plane distortions of the honeycomb lattice of \NIO\ involving correlations beyond that of the Ir-Ir near neighbor.

\end{document}